# Upper limit on the cross section for reactor antineutrinos changing $^{22}$Na decay rates

R.J. de Meijer[a,b,*],

a) Stichting EARTH, Weehorsterweg 2, 9321 XS, Peize, The Netherlands, rmeijer@geoneutrino.nl.
b) Dept.of Physics, University of the Western Cape, Private Bag X17, Bellville 7537, Republic of South Africa.

S.W. Steyn[c]

c) Koeberg Operating Unit, Eskom Holdings SOC Limited, Private Bag X10, Kernkrag 7440, Republic of South Africa, steyns@eskom.co.za.

Abstract

In this paper we present results of a long-term observation of the decay of $^{22}$Na in the presence of a nuclear fission reactor. The measurements were made outside the containment wall of and underneath the Koeberg nuclear power plant near Cape Town, South Africa. Antineutrino fluxes ranged from $\sim 5*10^{11}$ to $1.6*10^{13}$ cm$^{-2}$ s$^{-1}$ during this period.

We show that the coincidence summing technique provides a sensitive tool to measure a change in the total decay constant as well as the branching ratio between EC and $\beta^+$ decay of $^{22}$Na to the first excited state in $^{22}$Ne. We observe a relative change in count rate between reactor-ON and reactor-OFF equal to $(-0.51\pm 0.11)*10^{-4}$. After evaluating possible systematic uncertainties we conclude that the effect is either due to a hidden instrumental cause or due to an interaction between antineutrinos and the $^{22}$Na nucleus. An upper limit of $\sim 0.03$ barn has been deduced for observing any change in the decay rate of $^{22}$Na due to antineutrino interactions.

*) Corresponding author: Weehorsterweg 2, 9321 XS, Peize, the Netherlands, phone +31-505016654.

1. **Introduction.**

The search for new physics beyond models such as the Standard Model (SM) relies strongly on high precision measurements. This also holds for measurements involving neutrinos and antineutrinos. The most commonly applied and studied reaction for low-energy (<10 MeV) electron antineutrino detection is the inverse β-decay reaction where an electron antineutrino is captured by a free proton, resulting in the emission of a positron and a neutron. The overall cross section of this capture process is on the order of $10^{-43}$ cm$^2$. In this paper antineutrinos should be read as electron antineutrinos.

Nuclear power reactors are the strongest antineutrino sources accessible for experimentation. It was at a reactor and with the inverse β-decay reaction that Cowan and Reines (1956) first detected antineutrinos. It was also how various antineutrino properties were determined near the reactors in Bugey (Fr), Chooz (Fr) and at the KamLAND detector in Kamioka, Japan. There, detectors containing many tons of organic scintillation liquids provided the vast number of free protons required for a workable count rate.

The inverse β-decay reaction has the disadvantage that it has a Q-value of Q = -1.8 MeV. Since the antineutrino spectrum of reactors peaks below 0.5 MeV, a substantial portion of the antineutrino flux has insufficient energy for this interaction to take place. As already advocated by Weinberg (1962) for relic cosmic-neutrino detection, neutrino capture on β-decaying nuclei has a zero-energy threshold, and, in addition to the increase in the number of antineutrinos available for interactions, also expands the phase space. In fact studying antineutrino interactions with β$^+$-decaying nuclei at nuclear reactors provides an unique way to study antineutrino properties at very-low energies.

A few years ago several papers were published on a possible effect of solar neutrinos on nuclear β$^-$ decay (Jenkins et al., 2009, Jenkins and Fischbach, 2008). These papers were triggered after further analysis of earlier measurements of the β$^-$ decay of $^{32}$Si by Alburger et al. (1986) and of $^{226}$Ra and progeny, as well as γ-rays following β$^-$ decay of $^{152}$Eu (Siegert et al., 1998) revealed oscillations superimposed on the exponential decay. The oscillations in the decay rate have a magnitude of the order of $10^{-3}$, an oscillation period of one year, and extend over a period of several years. Jenkins et al. suggested that these oscillations may be attributed to the change in solar neutrino flux caused by the annual variation in the distance between the Sun and Earth. Several papers have challenged this hypothesis and attributed the effect to e.g. temperature changes inducing changes in solid angle (Semkow et al., 2009) or did not observe the effect (Norman et al., 2009, Bellotti et al., 2013, Kossert et al., 2014). Others (e.g. Siegert et al., 1998, Schrader, 2010) attribute the effect to discharges in the electronics caused by e.g. radon and thoron and their progeny.

Lindstrom et al. (2010, 2011) with a novel method, used the shape of a $^{198}$Au source to affect a change in possible self-induced activity resulting in a change in its half-life and found no significant effect at a maximum antineutrino flux of 3.5x10$^{11}$ cm$^{-2}$s$^{-1}$. Here an internal antineutrino flux was created by the β$^-$ decay of $^{198}$Au. The flux seen by the same mass of gold differs for a spherical and a rod shape source. Previously de Meijer et al.

(2011) reported on an experiment in the vault of the 2MW research reactor of the Delft University of Technology, the Netherlands. At a flux of $5 \times 10^{10}$ cm$^{-2}$s$^{-1}$ no significant effect ($\Delta\lambda/\lambda = (-1\pm1)*10^{-4}$ was found either; the corresponding upper limit for the cross section in this case was $3*10^{-23}$ cm$^{-2}$.

The motivation for the present experiment was to lower the upper limit of the cross section of antineutrinos modifying nuclear $\beta^+$-decay and thereby assess its potential as a means of monitoring the status of nuclear power reactors as well as possibly their fissile content. Moreover a significantly higher cross section than $10^{-43}$ cm$^{-2}$ would mean that the universe is not transparent to (anti)neutrinos and hence being a possible explanation for dark matter. It is further a more accessible testing ground for equipment needed if such experiments are to ever look for low-energy antineutrinos from the cosmic background as suggested by Weinberg (1962) as well as for dark matter experiments.

This paper first presents the methodology with a subsection on the experimental set-ups and histories, followed by a section outlining the analytical procedures describing the handling of the data and the way the data have been corrected for instrumental effects. In section 3 the results are presented. Section 4 deals with the possible biases caused by systematic uncertainties and errors due to the analytical procedures, pile-up and baseline restoring, the antineutrino flux estimate and the background. In section 5 the results will be discussed and conclusions will be presented.

## 2. Methods

2.1 Experimental set-ups and histories

Three sets of $^{22}$Na $\beta^+$-decay measurements were carried out at Koeberg Nuclear Power Station, ~30km northwest of Cape Town, South Africa in a few periods between 11 April 2011 and 9 February 2014. Scheduled and unscheduled power outages of the reactors were used to correlate count rate changes with changes of the antineutrino flux.

*2.1.1 First two sets of measurements*

For the first period ( April-June 2011) the set-up consisted of a 7.5cm diameter, 7.5 cm long cylindrical LaBr$_3$ crystal coupled to a photomultiplier tube (PMT), equipped with a pre-amplifier. The detector, the PMT and its pre-amplifier were obtained on-loan from the European Space Agency via the Delft University of Technology in the Netherlands. Standard NIM-bin electronics were used together with an ATOMKI MCA. Two 3kBq $^{22}$Na sources were placed in front of the detector. The set-up was shielded by a 7.5cm thick lead cylinder with a copper inner-lining. At both ends the cylinder was closed off by a 5 cm thick lead disk lined with copper on the inner side. Spectra, collected in 10 minute intervals, were stored in a laptop computer. They were then energy stabilised, quality inspected and grouped into one-hour spectra off-line.

This set-up was installed at a distance of ~23 m from the centre of the reactor core of unit#2. It was placed in a corridor near the containment wall. At this position the centre of unit#1 reactor core was ~66 m away. Here conditions were far from ideal with high noise levels, as well as high and variable temperatures. Except for a short interruption,

unit#1 ran at its full power of 0.9 GW$_e$. During the ramp-up of the unit#2 reactor, problems were encountered with the newly purchased main amplifier forcing the use of a replacement. Reactor-OFF measurements were therefore made with the initial amplifier while reactor-ON data were gathered using the replacement amplifier. On analysis of the data from the two periods it was concluded that the differences of $10^{-3}$ seen in count rate could not reliably be ascribed to a change in reactor status leaving a need for further measurements.

For the second set of measurements (February-August 2012) the LaBr$_3$ detector was no longer available necessitating the use of a NaI and a planar HPGe detector from iThemba LABS together with the lead shielding used in the first measurement. Both detectors were equipped with standard NIM-bin electronics and digitized with an ATOMKI MCA before stored into a laptop for off-line energy stabilisation and analysis. This set up was placed in the seismic vault of the reactor building well below ground level, directly under the reactor at a position about 16 m from the centre of the core of unit#1 corresponding to a distance of ~82m from the centre of unit#2. Here there is about 7m of reinforced concrete between the set up and the reactor core leaving a space of $2*3*2m^3$ between the concrete pillars, concrete floor and ceiling to work in. The weak ventilation in this large underground vault and the enormous mass of the surrounding concrete results in the temperature and humidity remaining relatively constant during the year. Each of the two detectors used a 3kBq $^{22}$Na source.

Due to limitations on the transport of LN$_2$, the HPGe detector was mainly used to assess the background as well as possible differences in the observed spectrum during both reactor-status ON and OFF. As no obvious differences in the background spectrum could be seen between reactor-ON and -OFF, the HPGe detector was disengaged. The NaI detector used had signs of deterioration from previous use, showing up as asymmetric peaks. Moreover the system showed pile-up in both pulse shape and peak position. The former was evident in the difference in peak shapes between background (BG) measurements (no source) and with source (Na+BG). One of the consequences was that neither a reliable background subtraction nor an off-line energy calibration could be applied. The data were not of sufficient quality to warrant further analysis and a third series of measurements was prepared.

*2.1.2 Third set of measurements*

The third set of measurements was conducted at the same location and position under the reactor as the second set. Despite the fact that the quality of the data from the first two sets of measurement was considered to be insufficient, the experience helped in the planning of the set-up of the third series of measurements as it became clear that a stable measurement environment is a prerequisite for long-term, high-quality data. A compact data-acquisition incorporated in the photo-multiplier tube (PMT) base has considerable advantages of e.g. recording various parameters such as the fine gain adjustment, PMT base temperature, and others, together with the spectra. Moreover on-line spectrum stabilisation is available which eliminated the need for less accurate offline spectrum drift corrections.

From analysing the first two measurements it also became clear that coincident summing would provide a way of experimentally separating events from the branches via β+ emission and via electron capture (EC). Based on these experiences, the set-up was redesigned for optimum sensitivity for coincident summing and making use of the advantages of a compact hard- and software acquisition system. For the third series the detector used was a 10.4 cm long, 10.4 cm diameter cylindrical NaI crystal with a 12mm diameter, 51mm deep well in which the source can be placed. The detector is optically coupled via a quartz light guide to a 76 mm diameter low-K, Electron Tube 9305KB PMT which is surrounded by a solid mu metal shield and equipped with a 14 pins connector. This detector set-up was manufactured by SCIONIX–Holland: model 102 BP102 / 3M-X. It plugged into an ITECH Venus base. The Venus is a universal digital multichannel analyser integrated into a photomultiplier base. It incorporates all the necessary electronics and was connected to a LENOVO laptop via the USB port. The on-line spectrum stabilisation and spectra storage was controlled by the INTERWINNER software provided with the Venus base. The detector and the PMT were housed inside a cylindrical lead shield with 7.5cm lead walls and an inner copper lining. At the top there are two 7.5cm thick, dove tailed sliding doors of lead; at the bottom there is an 85mm diameter hole through which the VENUS-base protrudes. Directly below the bottom of the vertical lead cylinder, at a distance of about 20cm, there are two 5cm thick lead discs that provide shielding against radiation from the floor.

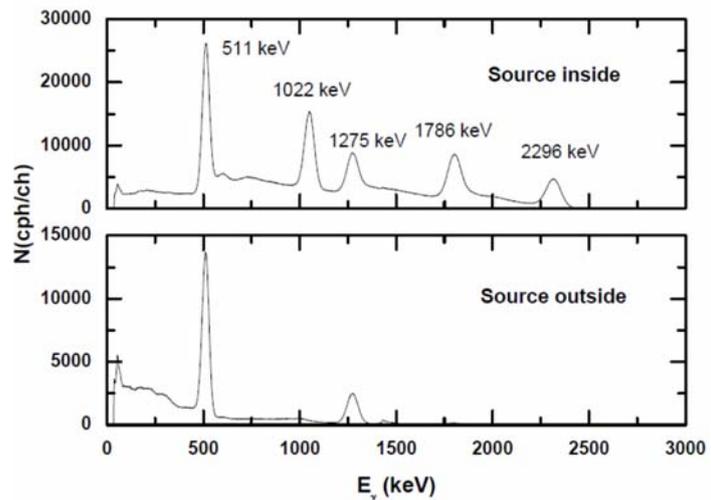

Figure 1.. Gamma-ray spectra with the NaI well detector and the $^{22}$Na source inside the well (top) and outside the detector (bottom).

The $^{22}$Na source is glued into a Perspex holder then wrapped in a 1mm thick Al foil and surrounded by ~3mm thick Pb foil. The Al and Pb foils absorb the positrons and the X-rays produced by the source. The source strength was estimated to be ~2kBq from the count rate at the start of the measurements.

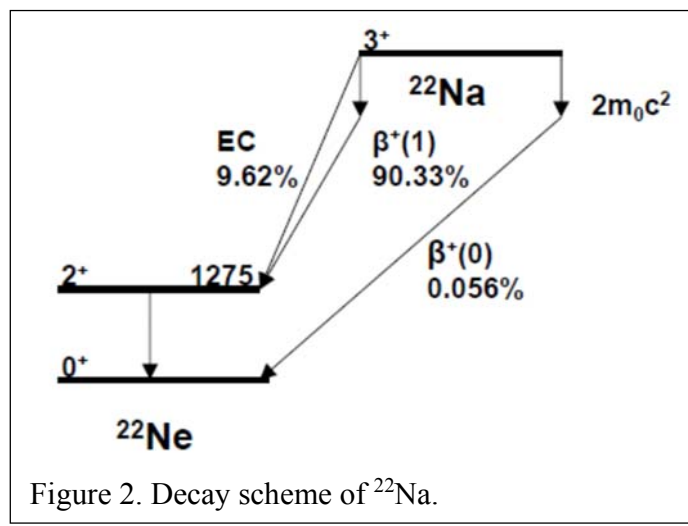

Figure 2. Decay scheme of $^{22}$Na.

Shown in Fig. 1 are spectra taken with the source inside and outside the well of the detector, which illustrates the effect of coincident summing in reducing the low-energy part of the spectrum and strengthening the high-energy part of the spectrum. The bottom panel shows the "traditional" spectrum of $^{22}$Na with the prominent 1275 and 511 keV peaks due to the de-excitation of the first excited state of $^{22}$Ne and one of the quanta from the positron annihilation, respectively (see Fig. 2). With the source inside the detector, additional peaks show up at 1022 (2*511) keV, 1786 (1275+511) keV and at 2297 (1275+2*511) keV. The shape of the continuum part of the spectra has changed due to the summing effects. (It should be noted that we use the term continuum for the part of the spectrum on which the peaks reside and that spectra taken without a source are named background.) This change in continuum shape complicates the continuum estimate and hence makes the resultant net peak count rates systematically less reliable. The comparison of the two spectra clearly indicates that all events (except for the background in this energy range) above the 1275 keV peak are due to coincident summing only and hence are exclusively related to β$^+$-decay.

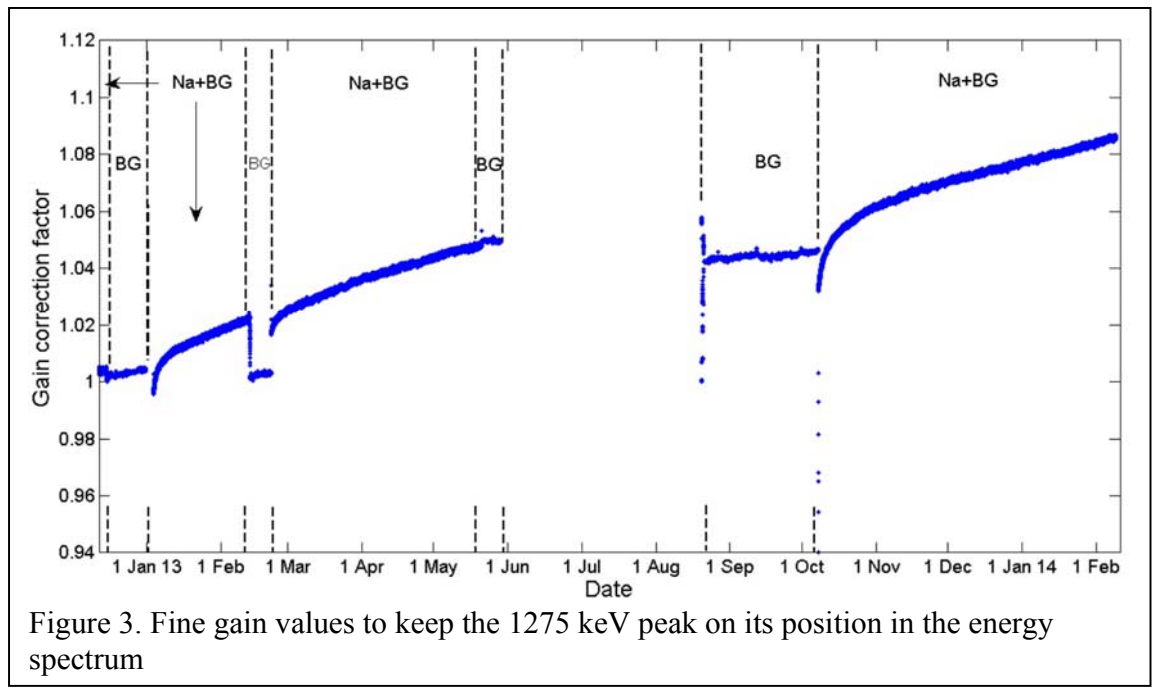

Figure 3. Fine gain values to keep the 1275 keV peak on its position in the energy spectrum

Data were acquired with this setup during two uninterrupted periods: 3 January-28 May 2013 and 20 August 2013-9 February 2014. During the first period a short BG measurement was conducted from 12 to 21 February. On 20 May a BG measurement was started, but on 28 May the system was inadvertently shut down, which went unnoticed until 20 August. On 20 August a long background measurement run was started, which continued until 8 October, on which date a new series of Na+BG measurements was started that lasted until 9 February 2014.

The online linear spectrum stabilisation was based on a gate around the 1275 keV peak. A fine-gain adjustment was automatically applied roughly every ten minutes. The adjustments were logged in a file for offline assessment. Figure 3 shows the applied fine-gain adjustments for the entire period. It is noticeable that during the BG measurements with a constant low count rate the applied fine-gain adjustment shows an only slight linear increase with time, but that, aside from the relatively large adjustments on start-up, the values during Na+BG measurements where the count rate decreases to good approximation linearly, the fine-gain adjustment linearly increase with time.

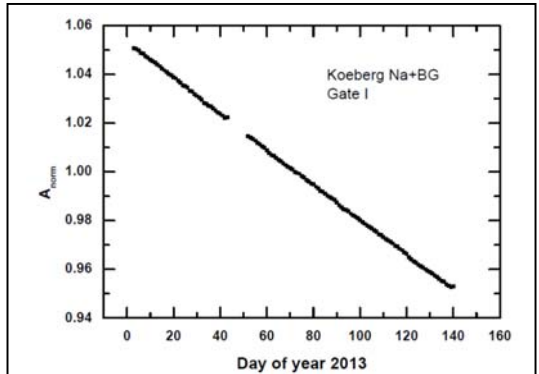

Figure 4. Normalised, to its average, count rate in Gate I as a function of time in the period 3 January-20 May 2013. No background subtraction nor decay correction have been applied.

This slow gain-drift phenomenon points towards a count rate effect such as insufficient pile-up correction or the baseline restoration not varying with the diminishing count rate. In both cases a decreasing count rate would reduce the pulse height. Consequently to keep the 1275 keV peak in the spectrum at a fixed channel position the fine-gain has to be increased as the count rate diminishes. If the adjustments are small the linear spectrum stabilisation seems to be sufficient, but during rapid changes of the value such as in the August-October 2014 period the lack of higher order terms in the stabilisation may cause transfer of counts from one region of interest (ROI) to another. Since the stabilisation centres around the 1275 keV peak, the lack of higher-order stabilisation will affect the high energy side more than the low energy side of the spectrum because the channel number is raised to a higher order (see also Sect. 4.2).

If a small change in the half-life of the decay is to be detected, a measurement over a period that is short compared to this half-life renders an insensitive method, as was also experienced by Lindstrom et al. (2011). To detect a relative effect of e.g. $10^{-4}$ the number of counts must be at least of the order of $10^8$. Such numbers are only achievable in long-term measurements and high count rates. But at the same token, small systematic changes may either enhance or mask such an effect present in the data and may not be traceable. Compared to absolute count rate measurements, measurements of count rate ratios are much less sensitive to systematic errors and uncertainties.

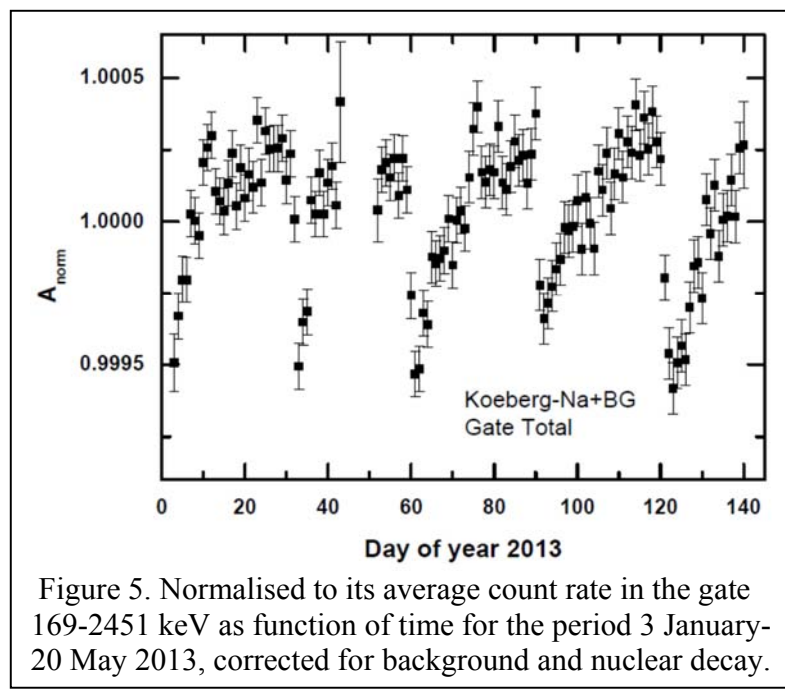

Figure 5. Normalised to its average count rate in the gate 169-2451 keV as function of time for the period 3 January-20 May 2013, corrected for background and nuclear decay.

2.2. Analytical procedures.

In the remainder of the paper energy ROIs will be set in the spectra and certain analytical procedures will be abbreviated. Table 1 presents an overview of the ROI settings and the abbreviations. In the figures presenting our measurements, error bars represent 1σ statistical uncertainties arising from propagating uncertainties based on normal distributions.

**Table 1. Settings of energy gates and abbreviations**

| ROI | Region of Interest |
|---|---|
| Gate Total | ROI in the γ-ray energy spectrum with 169<E<2451 keV. |
| Gate 1275 keV | ROI in the γ-ray energy spectrum with 1152<E<1351 keV. |
| Gate I | ROI in the γ-ray energy spectrum with 1353<E<1921 keV. |
| Gate II | ROI in the γ-ray energy spectrum with 1922<E<2451 keV. |
| Na+BG | Measurement with $^{22}$Na source. |
| DDC | Dip and decay corrected after subtraction of an average background (see text). |
| Pivoted | Count rate pattern fitted by a linear function and pivoted around its average value to a horizontal pattern. |
| ON/OFF | Reactor on power/ reactor shut down. |
| DOY | Day of the year 2013. |

In the decay of $^{22}$Na to $^{22}$Ne(1) electron capture results in a 1275 keV gamma ray only, whereas $β^+$-decay in addition has two 511keV photons. In the latter case all three quanta are emitted within the time resolution of the data-handling system and give rise to coincident summing. With coincident summing $β^+$-decay gives rise to stronger high-energy peaks and continuum in the spectrum. Therefore two very broad ROIs, Gate I and Gate II, were set to monitor the purely $β^+$-decay related events. Figure 4 shows the count rate evolution in one such a very broad ROI. The interruption starting around day 45 was for a background measurement to be made. The figure illustrates the dominance of the nuclear decay on the count rate.

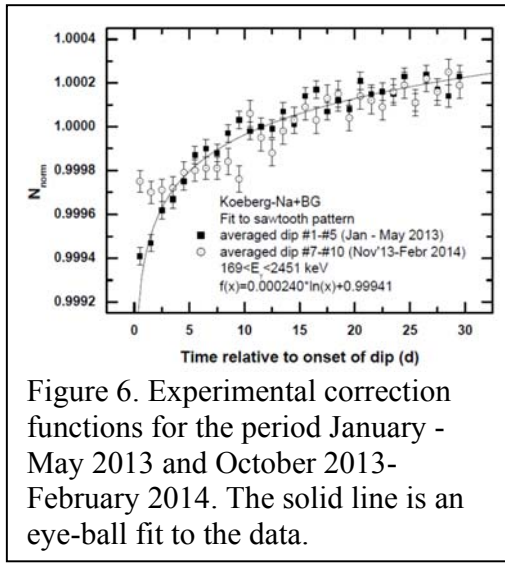

Figure 6. Experimental correction functions for the period January - May 2013 and October 2013- February 2014. The solid line is an eye-ball fit to the data.

To investigate possible weaker effects, the data were corrected for background and subsequently corrected for nuclear decay using the literature value for the decay constant ($λ=3.039*10^{-5}$ h$^{-1}$). An unexpected result was observed in all parts of the energy spectrum. Shown in Fig. 5 is the count rate, after background and decay corrections were applied, for a ROI set over almost the complete spectrum. A regular saw tooth-like pattern can be seen with an approximate periodicity of 30 days in the Na+BG data, but such a pattern was not observable in the much lower statistics spectra of the background measurements.

2.2.1 Dip corrections

Although the 30 day periodicity brought to mind a lunar influence, a detailed analysis revealed that the minima occurred at about noon on the first day of the month, irrespective of the length of the month, pointing to an artefact. Within statistics, an adjustment made to the date on the laptop in March 2014 corroborated this finding. If the dips are proportional to count rate then the minima would become shallower with time which is also observed. A software effect related to the management of time is now considered to be the most likely cause. The event scheduler on the laptop contains a programme LaunchRnR (" Lenovo uses this task to launch Rescue and Recovery for user to make Backup on local system"). The programme is started at 12:00pm on the first of the month. It is unclear how this programme affects the time management of the data acquisition.

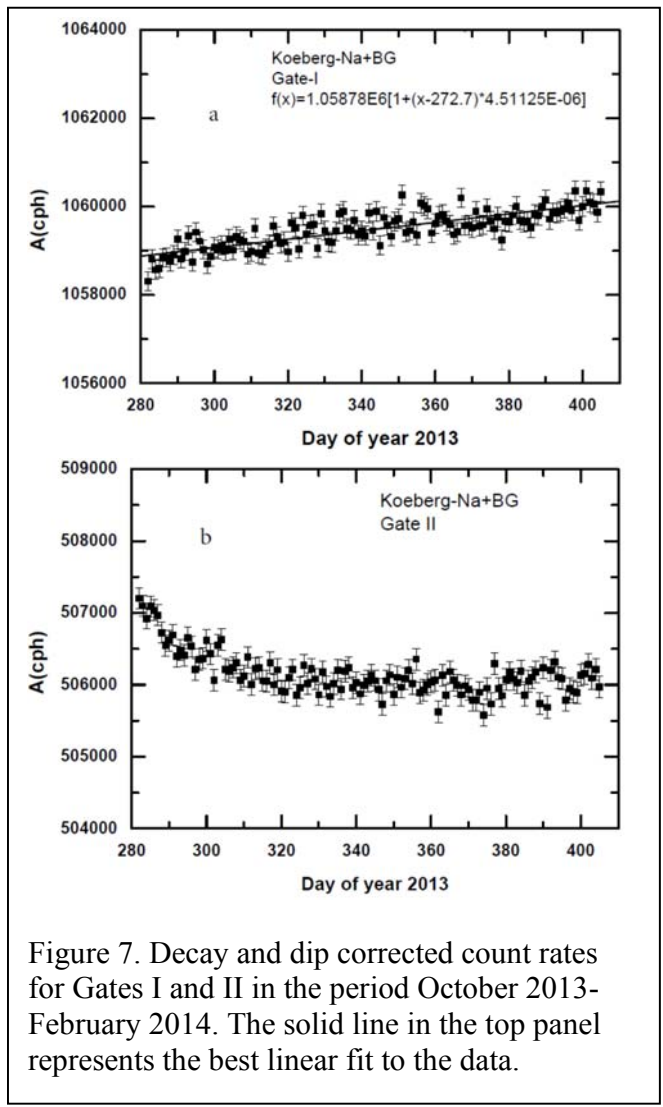

Figure 7. Decay and dip corrected count rates for Gates I and II in the period October 2013-February 2014. The solid line in the top panel represents the best linear fit to the data.

Under the assumption that the dip effect is an artefact, the data were separated into monthly periods and summed in order to obtain better statistics from which a correction function could be derived. The result is shown in Fig. 6. One notices that the correction function fits the data well except for the first two days of the November-February period where the minimum is not as deep and also shifted by one day. A correction calculated from the function was applied to all the data, except for the first two days in the second period where the values of the points plotted in Fig. 6 were used as the correction factor.

The effectiveness of the correction can be judged from Fig. 7 where the counts in Gates I and II are plotted after the decay and dip corrections have been applied for the period October 2013 to February 2014.

2.2.2 Sensitivity to online gain corrections.

Figure 7b (Gate II and furthest from the stabilisation peak at 1275 MeV) shows a gradually decreasing residual count rate during the first 30 days of the measurement, after which the count rate is almost constant. In Fig. 7a (Gate I and adjacent to the stabilisation peak) the residual count rate has a slight ($5*10^{-6}$ day$^{-1}$) linear increase with time. The solid line represents the best linear fit to the data and is used for a correction to the data by introducing a pivoting around the average of the count rate, which in this case occurs at day 340. The purpose of the pivoting is to obtain a horizontal count rate pattern for the entire reactor-ON-OFF-ON period. Subsequently for each of the reactor-ON and reactor-OFF periods weighted averages are determined separately to assess reactor-status effects (see Sect. 2.2.3).

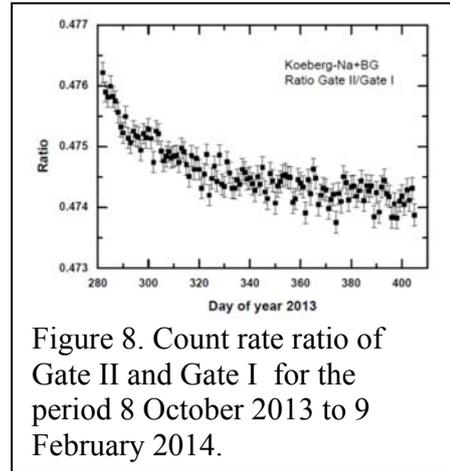

Figure 8. Count rate ratio of Gate II and Gate I for the period 8 October 2013 to 9 February 2014.

The decrease in count rate seen at the left of Fig. 7b can be understood from the fact that Gate II is further away from the stabilisation peak. The data are therefore more sensitive to any non-linearity in the gain, as well as the initial rapid gain changes. The starting date of the data in Fig. 7 is 8 October which falls in the 30 day period where a rapid change in fine-gain correction was still required, as can be seen from Fig. 3. Data as presented in Fig. 7b cannot be corrected by pivoting and have been excluded from further analysis.

The on-line stabilisation effect for Gate II as presented in Fig. 7b also effects e.g. the ratios involving Gate II as is shown in Fig. 8 for the ratio Gate II/ Gate I, which in principle should be independent of reactor status and insensitive to DDC.

2.2.3 Reactor-status effects

After the reactor is switched off, the fuel assemblies are left in the reactor for approximately a week to cool down and then are removed from the core and placed in the spent fuel pool. Decay antineutrinos however, continue to be emitted by the fuel. The gradual start-up of the reactor takes several days to almost a week before full power is reached. Vertical lines in Figs. 9-11 indicate the beginning and end of these transition periods on either side of the reactor-OFF period. The cool-off and start-up periods have been excluded from the weighted averages for the reactor-ON and -OFF periods.

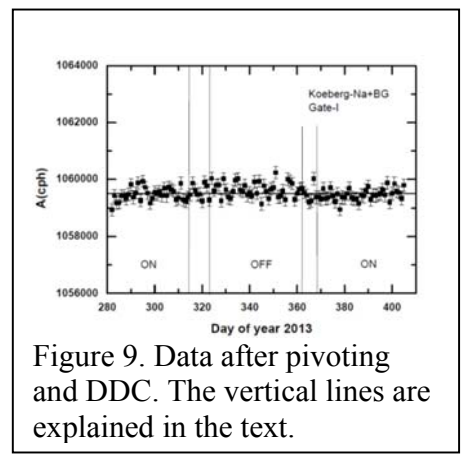

Figure 9. Data after pivoting and DDC. The vertical lines are explained in the text.

The result is presented in Fig. 9. To establish if the count rate is affected by the reactor status, the weighted averages of the count rate during the three periods are calculated from the corrected data. By taking the weighted average between the two ON periods, the effect of any remaining gradient in the data is reduced. With a linearly changing count rate curve and equally long measuring periods for the two ON periods the pivot is located at the centre of the OFF period. This implies that a small systematic error introduced by too much or too little rotation hardly affects the average value of the OFF-period, but it has an opposite effect on each of the ON-period weighted average values (bringing one ON value closer to the OFF-value and the other ON value further away) and their related $\chi^2$-values. In first order the opposite effects cancel by taking the average value of the two ON-period weighted averages.

The change in count rate relative to the reactor-OFF count rate: $\Delta A_{ON\text{-}OFF}/A_{OFF}$ is directly related to the change in decay constant, $\lambda$, and the associated cross section, $\sigma$, for the process leading to the change:

$$\frac{\Delta A_{ON-OFF}}{A_{OFF}} = \frac{\Delta \lambda}{\lambda}. \qquad (1),$$

and
$$\sigma = \frac{|\Delta \lambda|}{\Delta \phi_\nu} \qquad (2)$$

where $\Delta \phi_\nu$ is the difference in antineutrino flux between reactor-ON and reactor-OFF, under the assumption that all effects are due to antineutrino interactions only. A discussion on the systematic effects induced by the various approximations in the analytical procedures is presented in Sect. 4.1.

## 3. Results.

The analysis described above was carried out for a total of two ON-OFF-ON transitions of reactor unit#1: one in February-May 2013 and a second one in the period October 2013-February 2014. This resulted in the two data sets for Gate I and Gate 1275 keV plotted in Fig. 10. To reduce the statistical scatter of the results, 7 day averages are plotted.

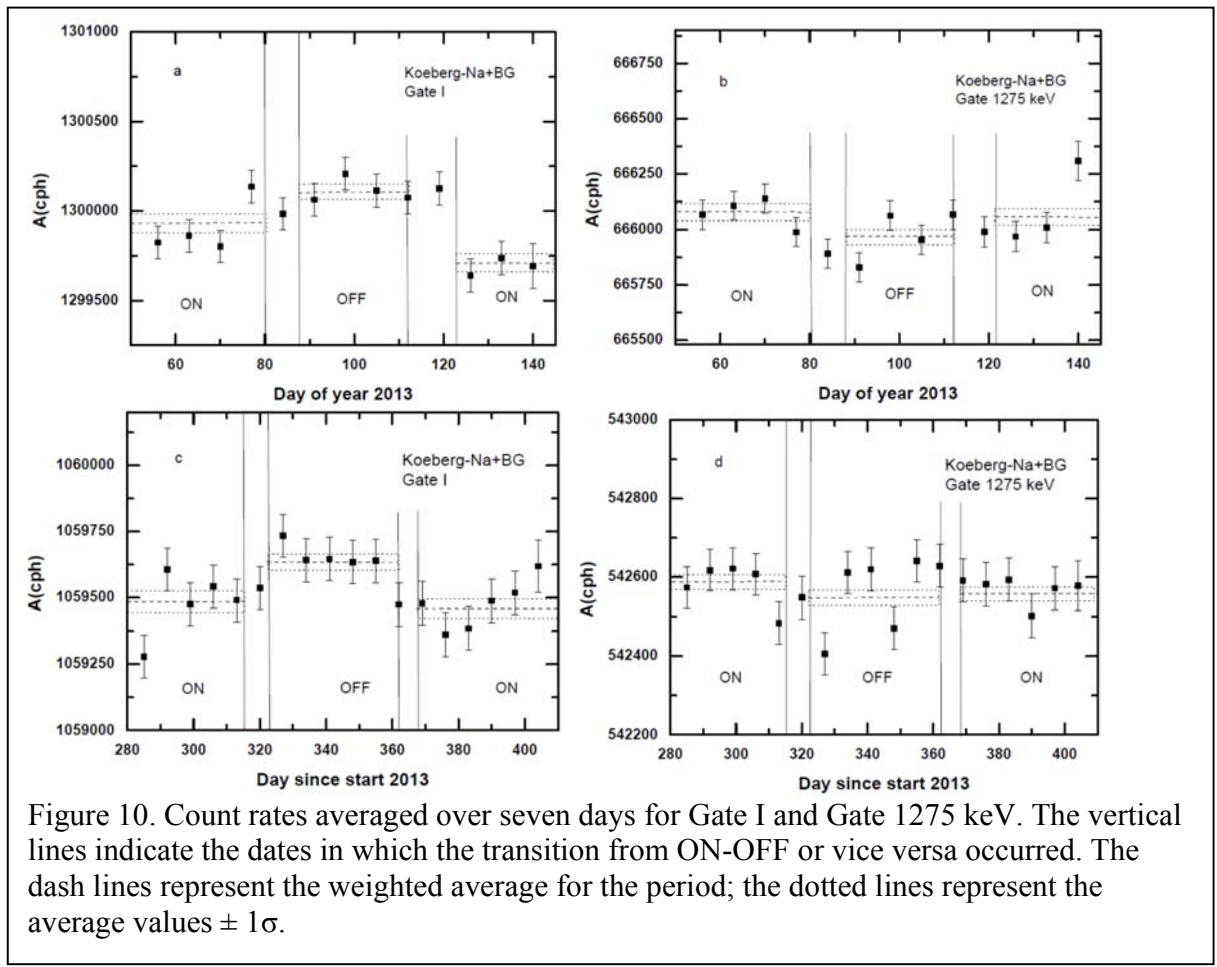

Figure 10. Count rates averaged over seven days for Gate I and Gate 1275 keV. The vertical lines indicate the dates in which the transition from ON-OFF or vice versa occurred. The dash lines represent the weighted average for the period; the dotted lines represent the average values ± 1σ.

From the data plotted in Fig. 10 one notices that the count rates decrease for Gate-I during reactor-ON whereas the count rates increase in Gate 1275 keV during the same period. Of interest is to note that while the count rate in Gate 1275 keV is affected by both $\beta^+$-decay and EC, whereas the count rate in Gate I is affected by $\beta^+$-decay only. One needs to caution against labelling this a physics effect as instrumental effects may still play a roll.

The fact that the count-rate changes, in the two ROIs as a function of reactor status, have opposite signs for Gate 1275 keV and Gate-I should have an effect on the count rate change for an almost total energy ROI Gate Total. Their count rates show a proper linear behaviour and hence may be analysed with the procedures described in Sect. 2. In Fig. 11 the count rate evolutions are presented for the periods February-May 2013 and October-February 2014. Both panels in the figure show a drop in count rate during reactor-ON periods. Included in Gate Total are the two ROIs under discussion plus $\beta^+$-decay related events such as the 511 and 1022 keV peaks and their associated continua.

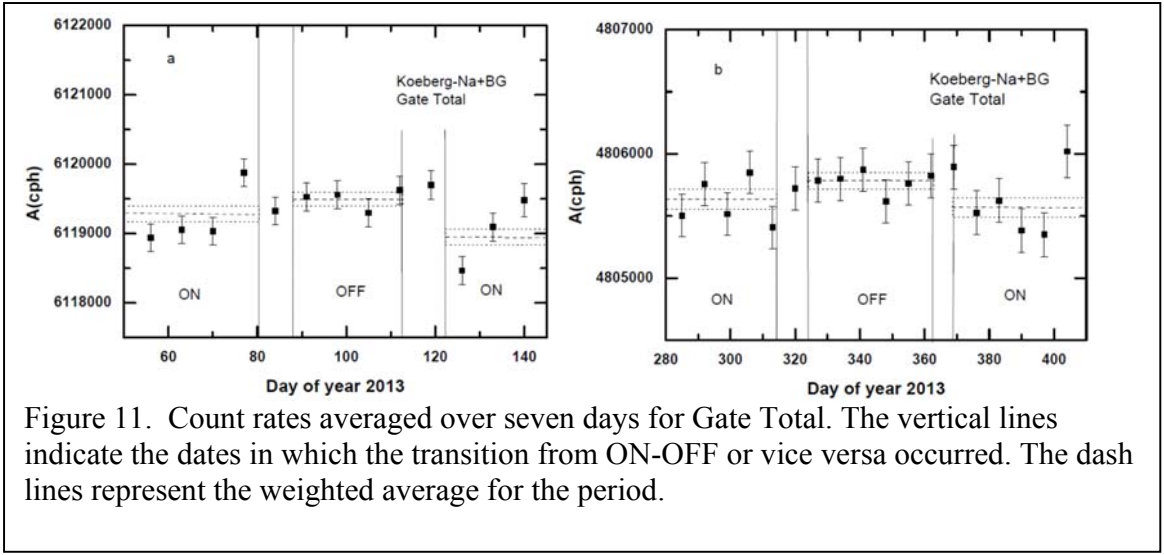

Figure 11. Count rates averaged over seven days for Gate Total. The vertical lines indicate the dates in which the transition from ON-OFF or vice versa occurred. The dash lines represent the weighted average for the period.

Table 2. Value of the weighted average count rate with corresponding $\chi^2$-values and the average difference between ON and OFF count rates. The uncertainties represent $1\sigma$ values.

| DOY 2013 | Status | $A_{total}$ (cph) (169-2451 keV) | $\chi^2$ | $A_{1275}$ (cph) (1152-1353 keV) | $\chi^2$ | $A_{Gate-I}$ (cph) (1353-1921 keV) | $\chi^2$ | $\left[\frac{\Delta A}{A_{OFF}}\right]*10^4$ total | $\left[\frac{\Delta A}{A_{OFF}}\right]*10^4$ 1275 | $\left[\frac{\Delta A}{A_{OFF}}\right]*10^4$ Gate I |
|---|---|---|---|---|---|---|---|---|---|---|
| 53-80 | ON | 6119280±110 | 1.41 | 666080±40 | 1.25 | 1299930±50 | 1.28 | | | |
| 87-112 | OFF | 6119490±100 | 1.11 | 665980±30 | 1.61 | 1300110±50 | 1.41 | -0.66±0.16 | 1.4±0.5 | -2.3±0.3 |
| 122-140 | ON | 6118950±110 | 1.79 | 666060±40 | 1.39 | 1299720±60 | 2.04 | | | |
| | | | | | | | | *-0.52±0.11* | *0.9±0.5* | *-1.9±0.4* |
| 282-315 | ON | 4805640±80 | 1.02 | 542590±30 | 1.32 | 1059480±40 | 0.97 | | | |
| 323-362 | OFF | 4805787±70 | 1.21 | 542550±20 | 1.33 | 1059640±30 | 1.87 | -0.39±0.16 | 0.4±0.5 | -1.6±0.3 |
| 368-389 | ON | 4805570±80 | 1.27 | 542560±30 | 0.26 | 1059460±40 | 1.15 | | | |

Table 2 presents the weighted averages of the count rates during a full reactor-status period (ON or OFF) used in Figs. 10 and 11. The $\chi^2$-values for a linear fit to the count rates. Also tabled is the average change of count rate during the two reactor-ON periods on either side of the reactor-OFF period, relative to the count rate during reactor-OFF period. The combined result of the two data sets is -0.52±0.11, +0.9±0.5 and -1.9±0.4, for Total, Gate 1275 keV and Gate-I, respectively, with external uncertainties quoted.

From the data listed in Table 2 it can be derived that for the combined ROIs, Gate 1275 keV and Gate I, the value of $\Delta A/A=(-0.9\pm0.3)*10^{-4}$ and $\Delta A/A=(-0.6\pm0.3)*10^{-4}$, for the two periods, respectively. These values indicate that the overall count rate changes with the values as obtained for Gate Total. In addition an exchange of count rate takes place between Gate 1275 keV and Gate-I. This exchange may carry information on changes in the partial decay constants.

**4. Systematic uncertainties and errors**

4.1. Analytical procedures

In Sect. 3 the results of the measurements have been described following the analytical procedures as outlined in Sect. 2. In these procedures only statistical uncertainties have been considered. The analysis results in possible effects of the order of $10^{-4}$. The procedures include corrections for saw-tooth shape dips, nuclear decay, and pivoting of data evolutions to allow the deduction of weighted averages. One may therefore question whether the accuracy of the results overrides the high precision.

The data set allows an assessment of the accuracy of the ΔA/A ratios. The ratio of the counts rates in the ROIs Gate 1275 keV and Gate-I, R, may be obtained in three ways:

1. As a direct ratio for every data point without a correction for the saw-tooth dips and the decay of the source. (No dip and decay correction: No-DDC).
2. The direct ratio for every data point after correction for the saw-tooth dips and the decay (DDC)
3. The ratio of the weighted averages for each of the reactor-ON/OFF periods. This method is indicated as RWA.

The first method contains the least assumptions and is based on the fact that common factors in numerator and denominator cancel in a ratio. These common factors include source strength and live time . In the second method the assumption is made that the dips are an instrumental effect, which may be corrected for by a correction function as described in Sect. 2.2.1 and illustrated in Fig. 6. The corrections for decay (source strength) and live time are the same for the corresponding data points and will drop out in the ratio. In the third method the effects of dip correction and decay correction and variations in live time end up in the weighted average, possibly after a linear fit to the data set of the three reactor-status periods and subsequently pivoted. This method contains the largest systematic uncertainty. The three methods are therefore listed in order of an increasing number of assumptions.

Table 3 presents a comparison of the results of the three methods during two periods in which the reactor status switched from ON-OFF-ON. For each of the periods there is no significant difference in the change in count rate ratio, ΔR, between the three analytical methods despite small differences between the corresponding R-values. The ΔR values for the two periods differ by a factor of two. Two possible explanations are: effects caused by the insufficient pile-up and/or base-line restoration were only partly compensated for by the linear spectrum stabilisation, and/or there was a change over time in the reactor-fuel composition. In the OFF-period in March-April 2013 no changes to the fuel composition were made, in contrast to the OFF-period in November-December 2013 when changes were made. From the results in Table 3 the explanation of the change in fuel composition is less likely because there is quite a change between the periods DOY 122-140 and DOY 282-315, whereas such a difference is absent between the periods DOY 282-315 and DOY 368-389.

**Table 3. Values of the count rate ratio R for three analytical procedures for two periods with a reactor-ON-OFF-ON sequence.**

| DOY 2013 | Status | $R_1$ No-DDC | $R_2$ DDC | $R_3$ RWA | $\Delta R_1 * 10^4$ | $\Delta R_2 * 10^4$ | $\Delta R_3 * 10^4$ |
|---|---|---|---|---|---|---|---|
| 53-80 | ON | 0.51503(3) | 0.51220(3) | 0.51239(4) | 1.6±0.5 | 1.7±0.4 | 1.5±0.5 |
| 87-112 | OFF | 0.51487(3) | 0.51204(3) | 0.51224(3) | *1.8±0.3* | *1.9±0.3* | *1.9±0.3* |
| 122-140 | ON | 0.51507(4) | 0.51225(4) | 0.51246(4) | 2.0±0.5 | 2.1±0.5 | 2.2±0.5 |
| | | | | | | | |
| 282-315 | ON | 0.51016(3) | 0.51200(3) | 0.51213(3) | 1.0±0.4 | 1.2±0.4 | 1.1±0.4 |
| 323-362 | OFF | 0.51006(3) | 0.51188(3) | 0.51202(3) | *0.9±0.4* | *1.0±0.4* | *1.0±0.3* |
| 368-389 | ON | 0.51014(3) | 0.51197(3) | 0.51211(3) | 0.8±0.4 | 0.9±0.4 | 1.0±0.4 |

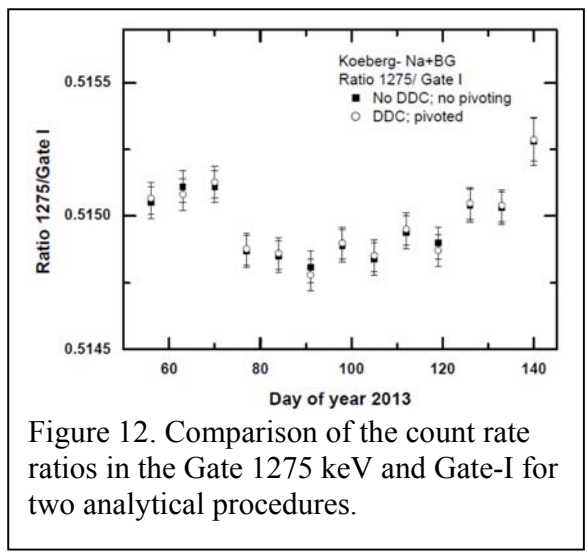

Figure 12. Comparison of the count rate ratios in the Gate 1275 keV and Gate-I for two analytical procedures.

Figure 12 presents the count rate ratios, averaged over seven days, for the analytical procedures in which corrections (DDC and pivoting) and no corrections (No-DDC, no pivoting) were applied. The DDC data were normalised to the No-DDC data, for comparison purposes. From Fig. 12 one may conclude that the differences between the two procedures are very small and confirm the results of Table 3 that the choice of the analytical procedure does not introduce a significant systematic uncertainty.

4.2 Pile-up and base-line restoring

Pulses of NaI scintillator detectors are known to have a long decay time. This makes measurements such as the present one sensitive to tail pulse pile up which, depending on the severity may affect the base-line restoration. Over time the decreasing count rate will result in an apparent pulse height decrease. The automatic spectrum stabilisation will keep the reference peak in the same channel by increasing the gain. As mentioned in Sect. 3.1.2 this effect is present in our measurements.

The average additional pulse height due to tail pulse pile up is a function of the average actual pulse height. For a well counter the coincident summing shifts the spectrum to higher channels and thereby increases the average additional pulse height. This results in a measurement such as the present one having a calibration that is more sensitive to count rate than would usually be the case.

The ITECH Venus-INTERWINNER combination supplies only a linear gain correction term. In time the lack of higher order terms in the energy calibration will redistribute counts over the various ROIs. This effect is likely the cause for the difference in ΔR and ΔA values for the two periods as listed in Tables 1 and 2. The effect is more noticeable over a longer period, but is not expected to significantly influence the ΔA/A or ΔR effects in a specific reactor-ON-OFF-ON sequence.

4.3 Antineutrino flux

After all corrections have been applied the remaining effect in the weighted average count rates for reactor-ON and -OFF under the assumption of an effect by antineutrinos on $^{22}$Na only, can be calculated as a cross section. This cross section and its accuracy depends on the estimate of the antineutrino flux. The fuel composition in the core is well balanced between assemblies at the beginning and end of their burn-up cycle. Table 4 presents the fuel composition at the beginning and end of the cycle for Koeberg, and the average number of neutrons and antineutrinos produced per fission as taken from Kessler (2011).

As may be concluded from Table 4, the contributions of the various fissioning nuclei changes by 30 to 50% during burn-up, but that the total flux drops by only 2% from the beginning to the end of a cycle. Taking an average burn-up for the entire fuel in the core, resulting in the assumption that the core produces antineutrinos homogeneously, the introduced inaccuracy in the total antineutrino flux is about 1%.

Next we consider the assembly configuration of the core to be cylindrical with diameter D=3.06 m and height H=3.66 m. This configuration may be further approximated by a sphere with its centre at 1.83 m from the bottom. Under these assumptions the $^{22}$Na source is at a distance of 16.1±0.5 m from the centre of the sphere. The power generated during reactor ON periods is 2.75±0.01 GW$_{th}$ for each unit. An assumption that 200 MeV is released in each fission event leads for this value of the power to an antineutrino flux from unit#1 of $1.48*10^{13}$ cm$^{-2}$s$^{-1}$ with an systematic uncertainty of about 15%, due to the uncertainties introduced in the estimate of the distance and the above assumptions on the configuration. The flux from unit#2 at the location of the source is $4.8*10^{11}$ cm$^{-2}$s$^{-1}$ and was constant during both reactor-ON-OFF-ON periods. Since unit#2 was at full power during the two reactor-ON-OFF-ON periods, unit#2 is irrelevant for the cross section calculation, which is based on the change in antineutrino flux between reactor-ON and reactor-OFF.

**Table 4. Neutron yield per nucleus, number of neutrons per fissioning nucleus, number of antineutrinos (anu) per fissioning nucleus, renormalized fission yield and the contribution to the total antineutrino flux per fission in the fuel at the beginning and end of the burn-up cycle.**

|  |  | n-yield | n/fission | anu/fission | fission yield | renorm. | anu |
|---|---|---|---|---|---|---|---|
| begin cycle | $^{235}$U | 0.67 | 2.5 | 5.58 | 0.27 | 0.65 | 3.61 |
|  | $^{238}$U | 0.08 | 1.7 | 6.69 | 0.05 | 0.11 | 0.76 |
|  | $^{239}$Pu | 0.25 | 2.9 | 5.09 | 0.09 | 0.21 | 1.06 |
|  | $^{241}$Pu |  |  | 5.89 | 0.01 | 0.03 | 0.18 |
|  | sum |  |  |  | 0.41 |  | 5.61 |
|  |  |  |  |  |  |  |  |
| end cycle | $^{235}$U | 0.42 | 2.5 | 5.58 | 0.17 | 0.41 | 2.27 |
|  | $^{238}$U | 0.08 | 1.7 | 6.69 | 0.05 | 0.11 | 0.76 |
|  | $^{239}$Pu | 0.50 | 2.9 | 5.09 | 0.17 | 0.42 | 2.12 |
|  | $^{241}$Pu |  |  | 5.89 | 0.02 | 0.06 | 0.35 |
|  | sum |  |  |  | 0.39 |  | 5.50 |

4.4 Background.

In the analysis an assumption is made that the BG is unaffected by the status of the reactor. Background measurements were made during both reactor-ON and reactor-OFF periods in the second experiment, but not during a reactor-status change. No significant difference was found, but there potentially still is a possibility that the BG count rate during reactor-ON may be slightly higher than during reactor-OFF. In the present analysis a BG count rate measurement was made during reactor-ON.

**Table 5. Differences in count rate between reactor-ON and -Off in terms of the Background count rate for three energy ROIs during the period DOY 39-139.**

| ROI | BG (cph) | Δ(ON-OFF) (cph) | Δ(ON-OFF) (%BG) |
|---|---|---|---|
| Gate Total | 57622±5 | -400±100 | -0.70±17 |
| Gate 1275 | 3422±2 | +90±30 | +2.7±0.9 |
| Gate I | 7130±2 | -300±40 | -4.2±0.6 |

Assume the background was lower for the reactor-OFF period, then for this period an overcorrection for background was made to the data and it would lead to the count rate during reactor –OFF being lower than during reactor-ON if no effect is present. In Table 5 for Gate Total, Gate 1275 and Gate I the Δ(ON-OFF) values are presented as a percentage of the BG count rate for the period DOY 53-139. The results in Table 5 indicate that the Δ(ON-OFF) effect cannot be described by an overall increase in BG count rate during reactor-ON. Gate Total, covering almost the entire energy range (169-2451 keV) and hence with the highest count rate in the BG spectrum shows the opposite sign to the expected value (see also Fig. 11). This observation excludes a higher background during reactor-ON than during reactor-OFF as a possible explanation for observed effects.

## 5. Discussion, conclusions and outlook.

The aim of the described investigations was to exclude antineutrino capture by $^{22}$Na as a potential tool for monitoring the status and possibly fissile content of nuclear power reactors. It was found that the count rate, corrected for the $^{22}$Na decay, shows during reactor-OFF an about $10^{-4}$ higher count rate than during reactor-ON for two ON-OFF-ON sequences of the reactor status. We come to two conclusions: either the decay of $^{22}$Na may be sensitive to the antineutrino flux changing with the status of a power reactor or the effect is due to a hidden systematic instrumental effect. As antineutrino detection is based on an indirect measurement via the decay of $^{22}$Na, systematic instrumental effects could not be ruled out. As discussed in Sect. 4 systematic effects due to various analytical procedures and/or pile-up effects do not explain the reactor-ON-OFF-ON effect, but does not exclude another instrumental mechanism. Whatever is the cause of the effect, there is an effect related to the ON-OFF status of the reactor, which may be applicable for nuclear reactor monitoring using compact detector systems.

These measurements took place outside the containment wall of and under the reactor, where no short term enhanced radiation fields are measured. This was confirmed by the BG measurements of the second series using an HPGe detector that shows no additional lines due to neutron activation. As explained in Sect. 4.4 a slightly higher background during reactor-ON has an opposite effect than measured.

Assuming the effect is due to antineutrinos, the cross section for the effect can be estimated from the $^{22}$Na source strength and the accepted antineutrino flux emanating from a power reactor of this type and size, the cross section of the process may be estimated from the result for Gate Total combined with the estimated change in antineutrino flux. The antineutrino flux was calculated assuming an estimated fissile fuel composition as well as the number of antineutrinos per fission as listed by Kessler (2011). Assuming an energy release of 200 MeV per fission, the change in flux of $1.0*10^{13}$ was calculated taking into account the distance to the reactor unit #1 (see Sect. 4.3) and ignoring flavour changes.

Assuming only antineutrinos affecting the count rate change in the Gate Total between reactor-ON and -OFF, the averaged change in decay constant over the two periods amounts to $\Delta\lambda/\lambda=\Delta A/A= (-0.52\pm0.11)*10^{-4}$. Applying for $^{22}$Na a decay constant of $\lambda=3.039*10^{-5}$ h$^{-1}$=$8.442*10^{-9}$ s$^{-1}$, according to Eq. 2, the cross section becomes $(2.9\pm0.6)*10^{-26}$ cm$^2$ with in addition a systematic uncertainty of about 15% (see Sect. 4.3) . A consequence of the assumption that the effect is due to antineutrinos only, is that the relative change in count rate of Gate-I, $(-1.9\pm0.4)*10^{-4}$, due to $\beta^+$-decay only, between reactor-ON and OFF is a measure for the change in the partial decay constant for $\beta^+$-decay. Together with the value for the total decay constant, this implies that $\Delta\lambda_{EC}/\lambda= (+1.4\pm0.4)*10^{-4}$. This result is consistent with a decrease in the decay constant of $^{22}$Na due to a reduction in the partial decay constant of $\beta^+$-decay and the related adjustments of the branching ratios of EC and $\beta^+$-decay to $^{22}$Ne(1).

A cross section of $10^{-26}$cm$^2$ exceeds the presently accepted calculated values on antineutrino interactions with $^{22}$Na by about 20 orders of magnitude (Cocco et al., 2007).

This enormous difference between this measurement and calculations indicates that either the measurements contain some subtle instrumental problem or the description of Cocco et al., 2007 is not adequate and the measurements indicate the presence of a yet unknown interaction between antineutrinos and the $^{22}$Na nucleus.

Large enhancement in cross section occur for many systems in atomic and nuclear physics at low relative energies. Strong final-state interactions in breakup reactions have been observed between p+p ($^2$He), p+t ($\alpha^*$) and $\alpha+\alpha$ ($^8$Be) ( see e.g. de Meijer and Kamermans, 1985). For the electro-weak interaction positronium is an example in which a positron and an electron form an unstable state. Positron interactions calculations with bound electrons in Li show an enhancement in cross section of more than ten orders of magnitude (Elbakry et al., 2013) at very low energies. If a similarity exists between electron-positron interactions and antineutrino-neutrino interactions a comparable reaction could take place between an incoming antineutrino and a neutrino in a bound proton leading to a strong enhancement in the cross section. In that case the antineutrino and neutrino annihilate and the decay of $^{22}$Na has a neutrinoless decay branch.

The above is very speculative and since a hidden instrumental effect cannot be ruled out the present results are taken as an upper limit for any process in which antineutrinos affect the $\beta^+$-decay of $^{22}$Na. The present value for the cross section is nevertheless about three orders of magnitude lower than the upper limit obtained in the Delft measurements (de Meijer et al, 2011) and four orders of magnitude lower than the solar neutrino effects on $\beta^-$-decay of $^{32}$Si as proposed by Jenkins et al. (2009). We may therefore repeat the previous conclusion (de Meijer et al, 2011) that either the hypothesis of Jenkins et al. is not true or that the effect of neutrinos on $\beta^-$-decay differs considerably from the effect of antineutrinos on the $\beta^+$-decay of at least $^{22}$Na.

This long term study with its 30 day periodic cycle highlights the dangers of accepting existing data from external measurements on face value and analysing them to the limit of the statistical precision while ignoring possible systematic errors. Also the sensitivity of data to small gain changes indicates that improvements to the stability of compact devices used for long term monitoring would greatly benefit from the reliability of data of measurements made with such devices. Adding on-line computing power such as in FPGAs several considerable improvements may be incorporated in PMT-base electronics, such as higher-order peak stabilisation during data taking and reduction of pile-up effects for systems that have a changing count rate due to decay. In addition, with the advent of cheap plentiful memory, more and more data from such long term measurements will become available.

5.1 Outlook
As stipulated a few times this investigation statistically shows a reactor status effect on the count rate of $\gamma$-rays emitted in the decay of $^{22}$Na, which may attributed to a hidden instrumental effect and/or an interaction of antineutrinos with $^{22}$Na nuclei. The latter would imply a cross section which is twenty orders of magnitude larger than the cross section of an antineutrino being captured by a free proton. About three orders of magnitude can be attributed to number of protons in a Na nucleus and the increased phase space for a zero-threshold reaction for the bound proton.

Such a large discrepancy between the measured and the expected cross sections demands further investigations. One option is repeating our measurements by others. For that reason we have described in this paper our measurements, analytical procedures and search for systematic uncertainties in quite some detail. In addition, we continue the measurements with the same detector, electronics and location with a $^{60}$Co ($\beta^-$)-source of similar strength, which should show no effect if an reactor status effect on $^{22}$Na is due to antineutrinos. Such a measurement is anticipated during the next scheduled outage of the reactor at Koeberg.

In parallel we are involved in the design, building and testing of an improved PMT-based data acquisition system. In this system a full-spectrum analysis technique will be used for spectrum stabilisation (see e.g. de Meijer, 1998).

In case the $^{60}$Co measurements indicate a physics effect we will apply the improved data acquisition system to measurements with $\beta^+$-sources like $^{44}$Ti and $^{26}$Al. These nuclei do not only contain more protons per nucleus than $^{22}$Na, but also have a much long half-lifetime. This implies that, aside of nuclear structure effects, for the same source strength more nuclei will be present and hence the reactor-status effects will have to be stronger. The outcome of such measurement will not only help us to understand the physics, but also brings a very-compact antineutrino detector for monitoring purposes onto the horizon.

**Acknowledgements**

The authors like to acknowledge the help of P. Dorenbos and F.G.A. Quarati for making the LaBr$_3$ available for the first experiment. We are very thankful the Koeberg Nuclear Power Station for providing us access to the facility. Especially we thank E. Welman and E. Reinier for their help in downloading and transmitting data from the set-up to us. We would like to thank Chr. Vermeulen, iThemba LABS, for manufacturing the $^{22}$Na source. The assistance of P. Papka during part of the second measurement is gratefully acknowledged.

The support in setting up the measurements, collecting data and discussing the results by F.D. Smit and M.W. van Rooy was highly appreciated.

We would also like to thank R. Lindsay, E.B. Norman and A. Zondervan for their comments and suggestions.

We are indebted to the Dutch IAEA Member State Support Programme for providing us with the NaI well counter and the associated electronics and software.

One of us (RdM) is grateful to the Dutch IAEA Member State Support Programme for contributing to the travel expenses and the hospitality at iThemba LABS.

The programme is part of the science and technology programme of Stichting EARTH in collaboration with its South African partners in a Memorandum of Understanding.